\documentclass[twocolumn,showpacs,preprintnumbers,amsmath,amssymb]{revtex4}
\usepackage{epsfig}
\begin{document}

\title{Current-Induced Magnetization Reversal in High Magnetic Fields in
Co/Cu/Co Nanopillars}
\author{B. \"{O}zyilmaz and A. D. Kent}
\address{Department of Physics, New York University, New York, NY 10003, USA}
\author{D. Monsma}
\address{Department of Physics, Harvard University, Cambridge, MA 02143, USA}
\author{J. Z. Sun, M. J. Rooks and R. H. Koch}
\address{IBM T. J. Watson Research Center, P.O. Box 218, Yorktown Heights, NY 10598, USA}
\date{12/16/02}

\begin{abstract}
Current-induced magnetization dynamics in Co/Cu/Co trilayer
nanopillars ($\sim$100 nm in diameter) have been studied
experimentally at low temperatures for large applied fields
perpendicular to the layers. At 4.2 K an abrupt and hysteretic
increase in resistance is observed at high current densities for
one polarity of the current, comparable to the giant
magnetoresistance (GMR) effect observed at low fields. A
micromagnetic model, that includes a spin-transfer torque,
suggests that the current induces a complete reversal of the thin
Co layer to alignment antiparallel to the applied field—---that
is, to a state of maximum magnetic energy.
\end{abstract}


\maketitle Angular momentum transfer in magnetic nanostructures
mediated by an electric current has become the subject of intense
research. Slonscewski and Berger first considered this
theoretically in 1996 \cite{Slonczewski1,Berger}. Its experimental
observation a few years later
\cite{Tsoi1,Sun1,Wegrowe,Tsoi2,Myers} has boosted efforts to
understand the influence of the conduction electron spin on the
magnetization dynamics of ferromagnetic nanostructures in the
presence of high electric currents. A spin current has been
demonstrated to switch the magnetization direction of a small
magnetic element. Reversible changes in multilayer resistance have
also been observed and associated with magnetic excitations or the
generation of spin waves. However, an in-depth understanding of
the relationship between these two phenomena--reversible
excitations of the magnetization and irreversible switching of the
magnetization--is still missing.

In the initial experiments a point contact was employed to inject
high current densities into a magnetic multilayer
\cite{Tsoi1,Tsoi2,Myers}. Subsequent experiments have concentrated
on the reduction of the lateral size of Co/Cu/Co trilayers to the
sub-micron scale, resulting in the fabrication of nanopillar
devices
\cite{Katine,Grollier1,Albert1,Grollier2,Albert2,Sun2,Sun3}. In
the point contact experiments the applied magnetic field was
oriented perpendicular to the thin film plane and was larger than
the film demagnetization fields (H$\geq$2 T). In this high field
regime a peak structure in the differential resistance (dV/dI) at
a critical current was interpreted as the onset of current induced
excitation of spin waves in which the current induced
spin-transfer torque leads to the uniform precession of the
magnetization \cite{Tsoi1,Tsoi2,Slonczewski2}. Spin-transfer
torque studies on pillar devices concentrated on magnetic fields
applied in the thin film plane. In the low-field regime, a
hysteretic jump in the differential resistance was observed; clear
evidence for current induced magnetization reversal of one of the
magnetic
layers\cite{Katine,Grollier1,Albert1,Grollier2,Albert2,Sun2,Sun3}.
Hence, the effect of the spin-polarized current on the
magnetization seemed to be quite distinct in the low and high
field regime. Experimentally, a current induced hysteretic
magnetization reversal was only observed at low \textit{in-plane}
fields in nanopillar devices.

In this letter we report detailed studies of current induced
dynamics of the magnetization in sub-micron size pillar devices at
high magnetic fields in the field-perpendicular to the plane
geometry. For sufficiently large currents of one polarity
hysteretic magnetic switching of the layers is observed at high
magnetic fields. In contrast to previous results in this geometry
with mechanical point contacts, these results cannot be understood
as small amplitude excitations of the magnetization. Micromagnetic
modeling suggests that spin-transfer torques induce precessional
states which evolve into a static state of antiparallel alignment
of the layers.

Sub-micron size pillar devices with the stack sequence of $|$3nm
Co$|$10nm Cu$|$12nm Co$|$300nm Cu$|$10nm Pt$|$ were fabricated by
thermal and electron-beam evaporation through a sub-micron stencil
mask \cite{Sun2}. TEM images show that this approach produces
pillar devices with steep side wall angles
($\sim7^{\circ}$)\cite{Sun3}. The thin Co-layer has a lower
coercivity and is the ``free'' layer in the device.
\begin{figure}
\begin{center}\includegraphics[width=8.5cm]{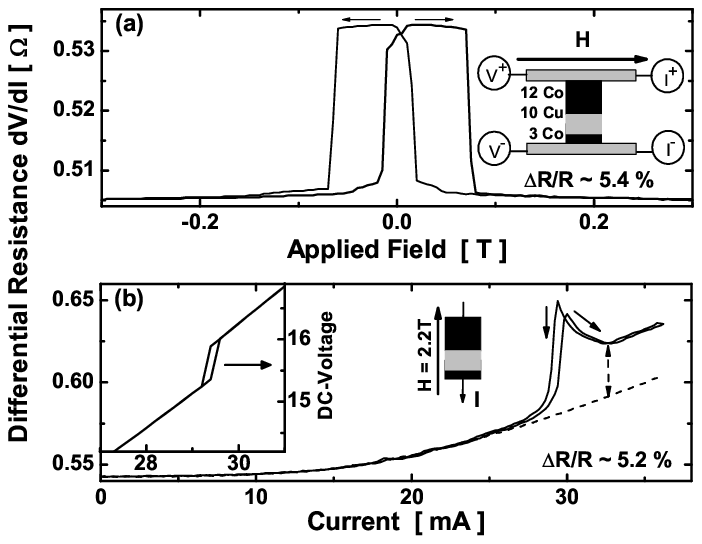}
\caption{(a) dV/dI versus H for fields oriented in the plane of
the thin-film layer and small bias currents. The inset shows the
layers sequence (in nm) and the 4-point measurement geometry. (b)
dV/dI versus I of the same device. The applied field is H=2.2 T
and oriented perpendicular to the film plane. At a critical
current a peak structure occurs in the differential resistance.
The dotted line shows dV/dI at negative currents, for which the
peak structure is absent. The inset shows the simultaneous
dc-voltage measurement.}
\end{center}
\end{figure}
The thick o-layer is the ``fixed'' layer in the device and acts to
setup a spin-polarized current in the intervening Cu layer. The
inset in Fig. 1(a) shows a sketch of the pillar device. Details of
fabrication and transport properties at low in-plane fields can be
found elsewhere \cite{Sun3}. Many junctions were studied
thoroughly as a function of current and applied magnetic field. In
this paper, we discuss representative data obtained on a sample of
lateral size 90 nm $\times$ 140 nm. All transport measurements
reported here were conducted at 4.2 K in a four-point measurement
configuration. The differential resistance dV/dI was measured by
means of phase-sensitive lock-in technique with a 100 $\mu$A
modulation current at $f$=800 Hz added to a dc bias current. The
dc voltage was recorded simultaneously. We define positive bias
such that the electrons flow from the free layer to the static
layer.

A typical magnetoresistance (MR) measurement of our pillar devices
at 4.2 K with an applied field in the plane of the thin films is
shown in Fig. 1(a). The device exhibits a clean transition between
a low resistance and high resistance state corresponding to
parallel (P) and antiparallel alignment (AP) of the magnetization
of the two Co-layers. At 4.2 K the MR of this sample is 5.4 \%.
Due to magnetostatic interactions between the layers, the high
resistance state is reached before the sign of H is
reversed\cite{Katine,Albert2}.

Fig. 1(b) shows dV/dI versus I with the applied field
perpendicular to the film. Here the applied field (H=2.2 T) is
larger than the demagnetization field of Co thin films (4$\pi$M
$\approx$ 1.5 T). The sweep to negative currents is shown as the
dashed line in the figure. From the data it is evident that the
abrupt change in resistance occurs only for one current direction.
In addition, the onset of this change in resistance is sharp and
takes place at a critical current, I$_{c}$
($j_{c}\approx2.3\times10^{8} A/cm^{2}$). Its characteristic
signature is a peak structure in the differential resistance
measurements. In the simultaneous dc-measurement [shown as inset
in Fig. 2(b)] this feature corresponds to a step-like increase of
the dc-voltage. The parabolic increase in the background
resistance for both directions of the current is due to increased
electron scattering at high current densities. However, thermal
effects cannot explain the abrupt and hysteretic change in device
resistance: these features are absent in pillar devices with only
a single Co-layer ($2 nm\leq t\leq 19 nm$) for current densities
up to $1.7\times10^{9} A/cm^{2}$ \cite{Barbaros}.

In earlier reports the peak structure in the differential
resistance of point contacts was attributed to the excitation of a
uniform precession of the free layer
\cite{Tsoi1,Tsoi2,Slonczewski2}. However, this picture cannot
explain our data. First, the resistance change is large. This can
be seen by either looking at the change of slope in the dc-voltage
at the critical current I$_{c}$ or by comparing the differential
resistance for I greater than I$_{c}$ with the differential
resistance at the same current value but with opposite polarity at
which the abrupt change in resistance is absent. Either comparison
shows a change in resistance of about 5 \%, similar to the GMR
value of the same device [Fig. 1(a)]. Therefore, an explanation in
terms of a small deviation of the free layer magnetization from
its parallel alignment with respect to the static layer is not
sufficient to explain the observed resistance change. With the
measurement of a resistance change comparable to the GMR effect,
it seems plausible to assume that even at high fields the spin
transfer effect can produce a full reversal of the
magnetization\cite{domain formation}. Second, the change in device
resistance is hysteretic, occurring at higher current density for
increasing current [Fig. 1(b)]. The excitations of spin waves
would decay rapidly on the time scale of such measurements and
thus appear reversible in such I-V measurements. Further, as we
show below, this interpretation is consistent with micromagnetic
modeling.

The experimental results of the magnetic field dependence of the
critical current is summarized in Fig. 2(a), in which the
differential resistance is plotted in a gray-scale. Here the
current is swept up from 20 mA to 36 mA while the magnetic field
is held constant for each current sweep. For subsequent current
sweeps the field is raised from 0.34 T to 2.7 T in steps of 5 mT.
In this gray-scale plot a critical current I$_{c}$ separates the
``applied field-current bias'' plane into two regions. Below the
critical current the device remains in its low resistance state,
in which both layers are in the parallel orientation. Above the
critical current, the magnetization of the device is in a higher
resistance state, in which the relative orientation of the
magnetization of the two layers deviates strongly from a parallel
configuration.

The critical current for decreasing current is shown as the dashed
line in this figure. Looking at the field dependence of the
hysteresis $\Delta$I$_{c}$ and the field dependence of the
critical current I$_{c}$, one can distinguish two regions. Above
1.4 T the critical current increases with applied magnetic field.
In this region also $\Delta$I$_{c}$ generally increases with
increasing field. (There is a deviation from this behavior between
2.0 T and 2.4 T, in which $\Delta$I$_{c}$ actually decreases. We
suspect this to be due to the onset of non-uniform excitations
that would reduce the hysteresis.) However, below 1.2 T the
critical current decreases linearly with applied magnetic field.
In this field region (1.1 T $\rightarrow$ 0.35 T), the hysteresis
is near the limit of our experimental resolution. The resistance
jumps for both regions are similar in magnitude, i.e. close to the
GMR value.
\begin{figure}
\begin{center}
\caption{(a) Magnetic field dependence of dV/dI versus I. The gray
scale represents the differential resistance of the junction.
Light color corresponds to high resistance and dark to low
resistance. The dashed line shows the position of the jump in
resistance for down-sweep of the bias current. (b) Micromagnetic
simulations of the experiment as described in the text.}
\end{center}
\end{figure}
To understand these results we consider the
Landau-Lifshitz-Gilbert (LLG) equations of motion for the free
layer including a spin-current induced torque predicted by
Slonczewski \cite{Slonczewski1,Slonczewski2}:

\begin{equation}\label{1}
\frac{d\hat{m}}{\gamma dt} =
-\hat{m}\times[\vec{H}_{eff}-\alpha\frac{d\hat{m}}{\gamma dt}]+
a_{I}\hat{m}\times(\hat{m}\times\hat{m}_{P})
\end{equation}

This zero-temperature, monodomain model is sufficient to
understand the basic physics of spin-transfer-induced magnetic
excitations \cite{Sun4}. Here $\hat{m}$ and $\hat{m}_{P}$ are
3D-unit vectors in the directions of the magnetization of the free
and fixed layer respectively. $\vec{H}_{eff}$ is the effective
field, $\vec{H}_{eff} = \vec{H}-4\pi M
(\hat{m}\cdot\hat{z})\hat{z}$, where $\vec{H}$ is the applied
external field, $\hat{z}$ is the film normal and $4\pi M=1.5 T$
for Co. $\gamma$ is the gyromagnetic ratio. The second term in the
brackets is the damping term and $\alpha$ is the Gilbert damping
parameter ($\alpha<<1$). The last term incorporates the
spin-transfer effects. The prefactor, $a_{I}$, depends on the
current, the spin-polarization
 of the current $\textit{P}$ and the angle between the
free and pinned magnetic layers $\Theta$, $a_{I} = \frac{\hbar
I}{eMV}g(P, \Theta)$ \cite{Slonczewski1}. Here \textit{g} is a
function of the polarization $\textit {P}$ that increases with
$\Theta$ and V is the volume of magnetic element. An increase in
$a_{I}$ with angle has been found in a number of different
approaches to modeling the spin-transfer effect
\cite{Bauer,Stiles,Waintal,Shpiro}.

In the large field regime ($H>4\pi M)$ the equation of motion can
be simplified. Initially both layers are parallel to the applied
field and the z-component of the free layer magnetization
satisfies (neglecting terms of order $\alpha^{2}$):

\begin{equation}\label{2}
\frac{dm_{z}}{\gamma dt} = (1-m_{z}^{2})[\alpha(H-4\pi M
m_{z})-a_{I}]
\end{equation}

From this equation it can be seen that an initial state m$_{z}$= 1
aligned with the applied field in the z direction will become
unstable when $a_{I}>\alpha (H - 4\pi Mm_{z})$. States for which
$|m_{z}|<1$ and $\frac{dm_{z}}{dt}=0$ correspond to precession of
the magnetization in the x-y plane at angular frequency of about
$\gamma (H-4\pi Mm_{z})$. Importantly, note that a solution with
magnetization antiparallel to the effective field (m$_{z}$=-1)
occurs for $a_{I}>\alpha (H - 4\pi M)$ and corresponds to a static
magnetization ($\frac{d\hat{m}}{dt}=0$). Thus the spin-torque, in
this case, leads to a state of maximum magnetic energy of the free
layer. Similar high energy states have also been observed in
numerical studies for the field in-plane geometry \cite{Sun4,Li}.
As $a_{I}$ increases with $\Theta$ the transition between a
precessing state with $|m_{z}|<1$ and $m_{z}=-1$ occurs rapidly
with increasing current and is hysteretic. For example, hysteresis
occurs when, $a_{I}>\alpha (H + 4\pi M)$ for
$\hat{m}\cdot\hat{m}_{P}=-1$ while at the same current
$a_{I}<\alpha (H - 4\pi M)$ when $\hat{m}\cdot\hat{m}_{P}=1$. In
addition, for increasing applied field, the transition occurs at
higher current and the width of the hysteresis increases.

Fig. 2(b) shows the result of integration of the eqn. 1 under the
conditions approximating the experiment ($P=0.4$ for Co and
$\alpha = 0.007$ (\cite{Schreiber})). The device resistance is
plotted on a gray scale versus I and H and is computed from the
angle between the fixed and free layers using the analytic
expression:
$R_{norm}=(R(\theta)-R(0))/(R(\pi)-R(0))=(1-\cos^{2}(\theta/2))/(1+\cos^{2}(\theta/2))$
\cite{Bauer,Shpiro,Wang}. For $H<4\pi M$ (below the horizontal
dashed line in the figure) the pinned layer magnetization tilts
into the plane and thus the pinned layer and effective field are
no longer collinear. In this case, we find precessional states of
the magnetization, with the projection of $\hat{m}$ on
$\hat{m}_{P}$ decreasing with increasing current. The average
resistance is plotted below this line. The dashed-dotted line
shows the transition to a lower resistance state for decreasing
currents, i.e., for states starting with the layers initially
anti-aligned.

Qualitatively there is a good correspondence between the
experimental data and the model. The model captures the general
features in the data, including the high field region of
increasing critical current with increasing applied field and the
low field region ($H<1.4T$) in which the critical current
increases with decreasing field. However, the model predicts
significantly lower critical currents (factor of $\sim5$) and more
hysteresis than that observed in the experiment. The latter is
perhaps not surprising as we have assumed single domain dynamics
in the model and likely the relaxation to the low energy magnetic
state occurs via non-uniform magnetic states of the free layer.

The strongest evidence for current induced magnetization reversal
at high fields comes from the experimental observation summarized
in Fig. 3. Here we show MR measurements at fixed current bias. For
low enough currents the spin-transfer torque cannot initiate
magnetization dynamics, independent of the magnitude of the
applied fields. An increase in magnetic field leads only to a more
parallel alignment of the two layers. An example for this case is
shown in Fig. 3(a). Here the bias current I is below a threshold
current I$_{t}$ $\approx$ 26.6 mA. The field dependence of the
device resistance changes dramatically, once the bias current
exceeds $I_{t}$ [Fig. 3(b)]. In this case, one can distinguish
between three distinct field regions. The boundaries $H_{1}$,
$H_{2}$ and $H_{3}$ of these three regions depend on the current
bias. In the first region ($H_{1}<H<H_{2}$) an increase of the
magnitude of the applied field leads again to a parallel
alignment. In the second region ($H_{2}<H<H_{3}$) precessional
states become possible and a gradual and \textit{reversible}
transition from parallel to antiparallel alignment takes place. At
the boundary between the second and third region ($H=H_{3}$) a
reversible but sharp transition from antiparallel back to parallel
configuration takes place. In the third region, where $H>H_{3}$,
the pillar device remains in the parallel configuration. A further
increase in current bias [Fig. 3(c)] shows that at high enough
currents the high field transition from antiparallel to parallel
alignment becomes \textit{hysteretic} whereas the low field
transition from parallel to antiparallel alignment remains
reversible. The width of the hysteresis depends on the current
bias and the polarity of the applied field. In general it
increases with increasing current bias.
\begin{figure}
\begin{center}\includegraphics[width=10cm]{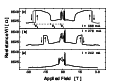}
\caption{Magnetoresistance measurements for a series of bias
currents I. (a) I is below a threshold current I$_{t}$. Beyond the
low-field regime H$>$H$_{1}$, an increase in field leads to a
parallel alignment of the layers. (b) I$>$I$_{t}$; current induced
torques lead toward antiparallel alignment in the field range
H$_{3}$$>$H$>$H$_{2}$. For H$>$H$_{3}$ the layers are forced back
to a parallel alignment. (c) Bias current is higher then in (b).
The abrupt transition becomes hysteretic, whereas the gradual
transition remains reversible.}
\end{center}
\end{figure}
This behavior is in sharp contrast to earlier reports
\cite{Tsoi1,Tsoi2,Katine}. In Ref. 8, when a high in-plane field
was applied to a pillar device, a large plateau in the
magnetoresistance with intermediate resistance $R_{int}$ value
($R_{AP}>R_{int}>R_{P}$)  was observed. The resistance plateau was
attributed to a precessing spin-wave state in-between P and AP
alignment.  From Fig. 3 it is clear that in the field
perpendicular geometry a large plateau in the magnetoresistance is
absent. In comparison to point contact experiments
\cite{Tsoi1,Tsoi2,Ji} high bias currents in pillar devices appear
to lead to a complete magnetization reversal even at high magnetic
fields. Our results thus suggest that the peak in dV/dI marks the
reversal of the free layer, not the onset of magnetization
dynamics. Of course, such DC measurements cannot rule out the
possibility of high field and current driven magnetic excitations.

In summary, we have presented evidence for current induced
reversal in Co/Cu/Co pillar devices at high magnetic fields. In
the field perpendicular geometry we have shown the existence of
two distinct field regions at which a transition between parallel
and antiparallel alignment takes place. In the low field region
the magnetization reverses gradually to an antiparallel alignment
through precessional states. In the high field region, at high
enough current bias, a hysteretic switching of the layers takes
place.

The authors gratefully acknowledge useful discussions with S.
Zhang, P. M. Levy and R. Kohn. This research was supported by an
NSF-FRG and by DARPA-ONR.


\begin{references}
\bibitem{domain formation} We associate this effect with a full reversal of of the free
layer as opposed to a break-up into domains \cite{Ji}, because the
free layer is much thinner than both the exchange ($\sim 10 nm$)
and spin-diffusion length ($\sim 60 nm$) in Co.
\end{references}
\end{document}